\documentclass[aps,prl,twocolumn,preprintnumbers,showpacs]{revtex4}

\def\gsim{\;\raise0.3ex\hbox{$>$\kern-0.75em\raise-1.1ex\hbox{$\sim$}}\;}
\def\lsim{\;\raise0.3ex\hbox{$<$\kern-0.75em\raise-1.1ex\hbox{$\sim$}}\;}
\newcommand{\be}{\begin{equation}}

\newcommand{\ee}{\end{equation}}
\newcommand{\bea}{\begin{eqnarray}}
\newcommand{\eea}{\end{eqnarray}}
\newcommand{\bt}{\begin{tabular}}
\newcommand{\et}{\end{tabular}}
\newcommand{\ba}{\begin{array}}
\newcommand{\ea}{\end{array}}
\newcommand{\ov}{\overline}

\begin{document}


\preprint{DSF-40/2003} \preprint{hep-ph/0311093}


\title{The effect of very low energy solar neutrinos on the MSW mechanism}



\author{S. Esposito}
\email{Salvatore.Esposito@na.infn.it}

\affiliation{Dipartimento di Scienze Fisiche, Universit\`{a} di
Napoli ``Federico II'' and I.N.F.N. Sezione di Napoli, Complesso
Universitario di Monte S. Angelo, Via Cinthia, I-80126 Napoli,
Italy}




\begin{abstract}
We study some implications on standard matter oscillations of
solar neutrinos induced by a background of extremely low energy
thermal neutrinos trapped inside the Sun by means of coherent
refractive interactions. Possible experimental tests are envisaged
and current data on solar neutrinos detected at Earth are briefly
discussed.
\end{abstract}


\maketitle



Experiments detecting neutrinos of energy up to several $MeV$
emitted by the Sun measure particle fluxes \cite{Exp}, \cite{SNO}
which are lower than those predicted by the Standard Solar Model
\cite{Bahcall}, apparently depending on neutrino energy. Recent
results obtained by the SNO collaboration \cite{SNO}, regarding
charged current as well as neutral current data, have convincingly
shown (with a high statistical significance level) that such a
deficit is explained in terms of transitions from $\nu_e$ into a
different (active) flavor state. Furthermore, compelling evidence
for atmospheric $\nu_{\mu}$ disappearance \cite{atmo},
independently checked by the K2K long-baseline experiment
\cite{K2K}, also points towards a natural explanation in terms of
neutrino flavor oscillations. However, while atmospheric neutrinos
would mainly change flavor during their travel in air
(practically, in vacuum), neutrinos emitted by the Sun could
experience coherent interactions with the solar matter, enhancing
the transition probability according to the MSW theory \cite{MSW}.
As observed, for example, in Ref. \cite{global}, the oscillation
scenario is nicely confirmed by a global analysis of solar,
atmospheric and reactor neutrino experiments, including KamLAND
recent results \cite{Kamland}. In particular, possible indications
in favor of solar matter effects in neutrino oscillation seem to
emerge from a combined analysis of SNO and KamLAND experimental
observations \cite{Fogli}. \\
\indent Solar neutrinos detected at Earth are produced deep inside
the Sun by the energy generating nuclear fusion reactions, and
their (observationally relevant) energy spectrum typically ranges
from hundreds of $keV$ to tens of $MeV$. However, lower energy
neutrinos are produced as well in the solar core by means of weak
charged and neutral current mediated thermal processes
\cite{Haxton}. The mean energy of such neutrinos is of the order
of the core temperature ($\sim 1.3 \, keV$), so that they account
for a negligible fraction of the energy loss ($\sim 0.001 \%$)
and, therefore, are potentially not relevant for the solar
evolution. Nevertheless, as pointed out in Ref. \cite{Haxton}, the
contribution of solar thermal neutrinos to the flux at earth is
dominant in the energy window above the cosmic background energies
($\gsim 10^{-2} \, eV$) and below the solar fusion and terrestrial
neutrino energy thresholds (around $5 \, keV$). Apart from
peculiar informations on the Sun properties carried by thermal
neutrinos, which would be useful for a further checking of the
Standard Solar Model, the detection of the existence of such
neutrinos will also translate into stringent kinematic limits on
the mass of muon or tau neutrinos. However, despite their
importance for Solar and Particle Physics, there is no obvious
experiment aimed to give a direct or indirect measurement of
thermal neutrinos, due to their low scattering cross section.

In this paper we will focus on extremely low energy neutrinos and
discuss an intriguing possibility for an indirect test of solar
thermal neutrinos, assuming that the higher energy ones (from
nuclear fusion reactions) experience flavor oscillations in the
Sun, according to the MSW theory.

\section{Neutrinos trapped in the Sun}

It is commonly believed that, due to their extremely small cross
section, very low energy neutrinos produced in a star freely
escape from it. This reasoning does not take into account,
however, coherent interactions with the particles in the plasma
which, as shown by Loeb \cite{Loeb}, become important at low
energy. In practice the refractive index of neutrinos with energy
$E$ in matter, of order $G_F N / E$ ($G_F$ is the Fermi coupling
constant and $N$ is the number density of particles in the plasma
interacting with neutrinos), approaches unity at very low
energies, so that a complete inner reflection takes place. As a
consequence trapped neutrinos exist inside the Sun (and, in
general, in every star). Smirnov and Vissani \cite{Vissani} have
pointed out the crucial role played by such neutrino sea in order
to solve the problem of an unphysically large value for the
self-energy of stars due to many-body long-range neutrino forces.

The weak interaction of thermal neutrinos with the particle in the
solar medium can be parameterized in terms of an effective
potential \cite{Raffelt}
\begin{equation}\label{1}
V_{\nu_e} \; = \; \pm \sqrt{2} \, G_F \left( N_e - \frac{1}{2} N_n
\right)
\end{equation}
for $\nu_e$ and
\begin{equation}\label{2}
V_{\nu_x} \; = \; \mp \frac{G_F}{\sqrt{2}} N_n
\end{equation}
for non-electron neutrinos ($x= \mu, \tau$). Here $N_e$ and $N_n$
are the electron and neutron number density of the plasma,
respectively; the upper sign refer to neutrinos, while the lower
one to antineutrinos. Effectively, the star can be viewed as a
potential well for neutrinos with depth given in Eq. (\ref{1}) or
(\ref{2}). Since, in the Sun, we typically have $N_e > N_n/2$,
$\nu_e$ and $\ov{\nu}_x$ experience a repulsive potential, while
an attractive one acts on $\ov{\nu}_e$ and $\nu_x$. As a result
the Sun expels $\nu_e$, $\ov{\nu}_\mu$ and $\ov{\nu}_\tau$, while
non-electron neutrinos and electron antineutrinos are trapped
inside it.

\section{Modified MSW effect}

For massive neutrinos, flavor eigenstates created by weak
interactions are, in general, linear superposition of the
propagating mass eigenstates, and the phenomenon of flavor
oscillations can take place \cite{Pontecorvo}. Assuming, for
simplicity, transitions between only two (active) flavor states
$\nu_e$ and $\nu_a$ ($a=\mu, \tau$), the evolution equation for
solar neutrinos reads as follows \cite{hamiltonian}:
\begin{equation}\label{3}
i \frac{d~}{dx} \left( \ba{c} \nu_e \\ \nu_a \ea \right) \; = \; H
\left( \ba{c} \nu_e \\ \nu_a \ea \right) ~,
\end{equation}
where the Hamiltonian matrix can be cast in the form
\begin{equation}\label{4}
H \;= \; H_0 \, + \, H_{\mathrm{int}}
\end{equation}
with
\bea
H_0 &=& \frac{\Delta m^2}{4E} \left( \ba{rr} - \cos 2
\theta & \sin 2 \theta \\ \sin 2 \theta & \cos 2 \theta \ea
\right) ~, \label{5}
\\
H_{\mathrm{int}} &=& \frac{1}{2} \left( \ba{rr} V & 0 \\ 0 & -V
\ea \right) ~. \label{6} \eea Here $E$ is the neutrino energy,
$\Delta m^2 = m_2^2 - m_1^2$ is the squared mass difference
between the two mass eigenstates and $\theta$ is the (vacuum)
mixing angle. The potential $V$ (depending, in general, on the
position $x$) is given by the difference between the $\nu_e$ and
$\nu_a$ interaction energies with the solar matter:
\begin{equation}\label{7}
V \; = \; V_{\nu_e} - V_{\nu_a} ~.
\end{equation}
In the ordinary MSW theory, assuming the (neutral) medium to be
composed of electrons, neutrons and protons, the potential in Eq.
(\ref{7}) depends only on the electron number density $N_e$
\cite{MSW}, \cite{Raffelt}:
\begin{equation}\label{8}
V \; = \; \sqrt{2} \, G_F \, N_e ~.
\end{equation}
However, as seen above, a sea of thermal neutrinos and/or
antineutrinos exists in the Sun which, in principle, influences
the propagation of fusion neutrinos escaping from the star and
detected at Earth. The interaction energies of $\nu_e$ and $\nu_a$
propagating in a medium composed also of thermal neutrinos with
number densities $N_{i}$ ($i = \nu_e, \nu_\mu, \nu_\tau,
\ov{\nu}_e, \ov{\nu}_\mu, \ov{\nu}_\tau$) has been calculated in
Ref. \cite{Raffelt} \footnote{Since $\nu_\mu$ and $\nu_\tau$ (and
their antiparticles) contribute in the same way to the potential
energies through neutral current interactions only, for simplicity
we set $N_{\nu_x} = N_{\nu_\mu} + N_{\nu_\tau}$ and similarly for
antineutrinos.} \bea V_{\nu_e} &=& \sqrt{2} \, G_F \left\{ N_e -
\frac{1}{2} N_n +
2 \left( N_{\nu_e} - N_{\ov{\nu}_e} \right) + \right. \nonumber \\
&~& \left. \vphantom{\frac{1}{2}}
\left( N_{\nu_x} - N_{\ov{\nu}_x} \right) \right\}  \label{9} \\
V_{\nu_a} &=& \sqrt{2} \, G_F \left\{ - \frac{1}{2} N_n + \left(
N_{\nu_e} - N_{\ov{\nu}_e} \right) + \right. \nonumber \\
&~& \left.  \vphantom{\frac{1}{2}} 2 \left( N_{\nu_x} -
N_{\ov{\nu}_x} \right) \right\} \label{10} ~. \eea As pointed out
above, the solar plasma acts through a repulsive potential on
thermal $\nu_e$ and $\ov{\nu}_x$, so that for simplicity we
definitively assume that $N_{\nu_e}=N_{\ov{\nu}_x}=0$ in the Sun.
The difference in Eq. (\ref{7}) then takes the form:
\begin{equation}\label{11}
V \; = \; \sqrt{2} \, G_F \, \alpha \, N_e
\end{equation}
with
\begin{equation}\label{12}
\alpha \; = \; 1 \, - \, \frac{N_{\ov{\nu}_e} + N_{\nu_x}}{N_e} ~.
\end{equation}
The effect of the thermal neutrino sea on the MSW mechanism is
thus simply parameterized in terms of the quantity $\alpha$ in Eq.
(\ref{12}): for $\alpha=1$ we recover the ordinary MSW theory.
Note that since the trapping of thermal $\ov{\nu}_e$ and $\nu_x$
is ruled by coherent interactions with the electrons (and
neutrons) in the plasma (see Eqs. (\ref{1}) and (\ref{2})), it is
natural to expect that, at a rough approximation, the parameter
$\alpha$ is nearly constant through the Sun. In this case the
structure of the evolution equation in (\ref{3}) does not change
(no $x$-dependent term appears) and the oscillation probability
changes only for a multiplicative constant factor in $V$.

\section{Theoretical predictions for the neutrino sea}

Very low energy neutrinos are created in stars by a number of
electroweak processes. Usually one can think about URCA processes,
neutrino pair bremsstrahlung or even production induced by
many-body long-range forces \cite{Vissani}. To our best knowledge,
the most complete study present in the literature regarding the
production od sub-$keV$ neutrinos has been performed in Ref.
\cite{Haxton}. Here the authors show that the relevant processes
in the Sun are: 1) Compton production; 2) intermediate plasmon
pole contribution to the Compton process; 3) transverse plasmon
decay into neutrino-antineutrino pairs; 4) decay of thermally
populated excited nuclear states into neutrino-antineutrino pairs
by means of $Z^0$ emission; 5) pair production in electron
transitions from free to bound atomic states. It is found that the
pole contribution to the Compton process dominates the production
of low energy ($<5 \, keV$) $\nu_e$ and $\ov{\nu}_e$ while, for
non-electron neutrinos, the dominant channel below $2 \, keV$ is
the free-bound process.

Let us now assume, as above, that extremely low energy
$\ov{\nu}_e, \nu_\mu, \nu_\tau$ are completely trapped inside the
Sun. A rough estimate of the number densities of such neutrinos,
relevant for the evaluation of the parameter $\alpha$ in Eq.
(\ref{12}), has been previously given in \cite{Vissani}. Assuming
thermodynamical equilibrium and strong fermion degeneration for
the neutrino sea, the number density of the particles in the bath
is $N_\nu \sim \mu^3 \sim V^3$, where $\mu$ is the neutrino
chemical potential which would be of the same order of magnitude
of the potential in Eq. (\ref{1}) or (\ref{2}). In this case the
resulting $N_\nu$ is exceedingly small, of the order of $10^{-28}
\, cm^{-3}$. However, in this estimate it has not been taken into
account that the production reactions (for example those
considered in \cite{Haxton}) render neutrino trapping a {\it
non-equilibrium} phenomenon. Then it seems natural to assume that
the number densities of trapped very low energy neutrinos is much
greater than the corresponding equilibrium values \footnote{Note
that if the sea is far from thermodynamical equilibrium, the Pauli
blocking mechanism could not give a stringent limit on the number
of neutrinos in the bath. In fact the Pauli principle forbids the
filling of one-particle states with more than one neutrino, but
the distribution with energy of these states is not related to the
Fermi distribution for non-equilibrium configurations. In
principle we can then think to peculiar distributions with many
states at very low energies and few states at higher energies,
where neutrinos are no longer trapped by refraction.}, as for
example happens in the early Universe plasma during primordial
nucleosynthesis \cite{BBN}. Therefore, since the Standard Solar
Model in its present form explains well the observed properties of
the Sun \cite{Bahcall}, we have only to ask that the energy
density of trapped neutrinos is much lower than the energy density
of the solar plasma of electrons, protons and neutrons. in order
to not affect solar dynamics. To obtain a rough limit on $N_\nu$,
let us assume for simplicity that the energy density $\rho$ of the
solar plasma is contributed only by electrons at a temperature of
order $1 \, keV$: $\rho_e \sim N_e \cdot <E_e> \; \sim N_e \cdot T
\sim 10^{14} eV^4$. By taking the typical energy of trapped
neutrinos to be of the order of the potential in Eq. (\ref{1}) or
(\ref{2}) (as also assumed in \cite{Loeb} and \cite{Vissani}), and
requiring that the energy density of the neutrino sea $\rho_\nu
\sim N_\nu \cdot E_\nu$ is much lower than $\rho_e$, we get the
very weak limit $N_\nu \ll 10^{41} \, cm^{-3}$. In practice, using
the above arguments, even if the number density ratio in Eq.
(\ref{12}) approaches the face value 1, the energy density of
trapped neutrinos would account for only $10^{-2} \, eV^4$ (i.e.
about 14 orders of magnitude less than electrons), and solar
evolution is completely unaffected.

\section{Experimental tests}

The indirect detection of the neutrino sea in the Sun can be
carried out by studying solar neutrino oscillations and measuring
the parameter $\alpha$ in Eq. (\ref{11}). The effect considered
here depends, of course, on the number density of the particles in
the sea, and any observed deviation from the standard MSW
prediction ($\alpha=1$) in the direction underlying Eq. (\ref{11})
($\alpha<1$) should be ascribed to the presence of a neutrino sea
in the Sun.

The feasibility of such kind of analysis, in the near future, is
well outlined in a recent paper by Fogli et al. \cite{Fogli}. Here
the authors perform global fits of current solar neutrino data
together with CHOOZ reactor observations \cite{CHOOZ} and recent
KamLAND results \cite{Kamland}, obtaining indications in favor of
neutrino matter oscillations in the Sun. This is achieved by means
of a $\chi^2$-analysis for the parameter $\alpha$ actually
appearing in our Eq. (\ref{11}): a value $\alpha=0$ would indicate
no matter effects (vacuum oscillations), while $\alpha=1$ is an
evidence for the (standard) MSW theory. However, in the light of
the mechanism proposed here, observed values for $\alpha$
different from 1 would give strong evidence in favor of the
existence of the solar neutrino sea.

As a result, with current data, the authors of Ref. \cite{Fogli}
find a best fit value for $\alpha$ close to unity (slightly
greater than 1) with an overall $\pm 3 \sigma$ range spanning
about three orders of magnitude (approximately $10^{-1} \lsim
\alpha \lsim 10^2$). While this allowed range is rather large, and
no firm conclusion can be reached neither on general matter
effects in solar neutrino oscillations nor on the presence of
effects induced by a neutrino sea, nevertheless it is interesting
to observe that simulated data for the KamLAND experiment with
increased (a factor 10) statistics would exclude values for
$\alpha$ greater than 1 {\it but not} those corresponding to
$\alpha <1$. Since a value for $\alpha$ {\it lower} than unity is
a peculiar prediction of the mechanism considered here (see Eq.
(\ref{12}), if such a poor indication will be confirmed by future
experiments, it would give a strong evidence for the sea of very
low energy neutrinos in the Sun.

\begin{acknowledgments}
The author is grateful to Drs. A. De Candia and G. Imbriani for
valuable discussions.
\end{acknowledgments}

\end{document}